# Evaluating the Impact of Lean and Green Practices on Operational Performance: A Real Data-Driven Simulation Case Study

Farah Altarazi, PhD

The State University of New York at Binghamton

New York, USA

faltara1@binghamton.edu

## ABSTRACT

Global market-driven forces and customer needs are continuously changing. In the past, profitability and efficiency were the primary objectives of most companies. However, in recent decades, sustainable performance has emerged as a new competitive advantage. Companies have been compelled to adopt a concept that combines these evolving global interests with traditional goals, resulting in the innovation of the lean and green approach.

In this study, a research methodology that includes system analysis and modeling procedures to apply the lean and green concept, combined with a new evaluation metric, the Overall Environmental Equipment Effectiveness (OEEE) was used to investigate the effects of adopting lean and green practices on overall performance.

A simulation model and energy value stream mapping were implemented, and the OEEE value was calculated to assess the current performance in terms of quality, availability, productivity, and sustainability. The current state production lead time was 329.1 minutes per batch, and the OEEE value was 13.1%. This result indicates existing issues in performance and sustainability, suggesting that improvement efforts should focus on enhancing these two aspects to increase the overall OEEE value.

Several improvement scenarios were proposed, including combining and rearranging the inspection workstations as the first scenario, and using UV lighting for drying purposes at the framing workstation as the second. After applying these improvements, both scenarios showed increased OEEE values and reduced lead times compared to the current state. In the first scenario, the lead time decreased to 158.23 minutes, and the OEEE increased to 35%. In the second scenario, the lead time was reduced to 292 minutes, with the OEEE increasing to 24%.

These findings demonstrate that applying lean and green practices can positively impact overall performance and significantly reduce production lead time.

**Problem Definition and General Overview**

In recent years, rapid industrial development increases the competition between companies in terms of price, cost, quality, and effectiveness where customers have become more demanding and more versatile, companies now don't have a choice; they have to maintain their high-quality products with minimum costs (Engin et al., 2019). Meanwhile, other challenges have arisen, governmental and social bodies are demanding companies to reduce their environmental impacts and wastes, and this is why we find a lot of companies are seeking to adopt green practices such as ISO 14001 regulations to maintain an effective environmental management system (Jiang & Bansal, 2003).

One of the most popular principles that have been suggested recently to encounter these challenges is the lean and green practices (Garza-Reyes, 2015). The implementation of applying lean and green principles showed significant environmental wastes reduction and saving costs, which encouraged a lot of companies to adopt these principles to improve their overall performance.

The lean management concept has been initially proposed by Toyota Production Systems, with the aim of identifying non-value-added activities-these activities which the customer will not be willing to pay for it – in addition to propose a different method to eliminate these activities in order to reduce costs and improve performance (Thanki et al., 2016). The application of lean management concept has been proved in many small and large factories its effectiveness in minimizing the wastes including transportation, inventory, motion, waiting, over-processing, overproduction, and defects (Atieh et al., 2016).

Green production is known as "The application of environmentally and socially sensitive practices to reduce the negative impact of manufacturing activities while, at the same time, harmonizing the pursuit of economic benefits" (Baines et al., 2012). Green production can be achieved in many ways and at different stages, for example, changing the type of raw materials, or lowering the consumption of energy during the production process can help to make the production system greener (Maruthi& Rashmi, 2015).

The intersection of both principles is known as lean and green practices, or Eco-Efficiency concept which can be defined as integrating different practices to improve operational efficiency and sustainability continuously (Abreu et al., 2017). Different lean and green tools have been developed in order to help in integrating this concept such as energy value stream mapping (eVSM) which will be used in this study along with simulation-based system and performance evaluation calculations.

The Overall Environmental Equipment Effectiveness (OEEE) has been chosen as an indicator to evaluate the adoption of lean and green practices. Overall Equipment Effectiveness (OEE) is a known tool to measure manufacturing productivity which includes the availability, quality, and performance of each production step. Moreover, (OEEE) incorporates the concept of sustainability based on the calculated environmental impact of each production step (Domingo et al., 2015).

Sproedt et al. (2015) has suggested a research methodology that involved the use of simulation and energy value stream mapping (eVSM) to evaluate the production system performance. in this study the concept of OEEE that described previously has been added as a performance evaluation indicator of the current state and to evaluate the future improvements.

**Literature Review**

This section represents the existing literature related to the implantation of the lean and green practices. An overview will be presented to show how the concept of lean and green practices and tools has developed, starting from the implantation of lean, green separately and how each practice affects the overall performance, and then the relation between them, and finally how the concept of lean and green has developed to become one entity to use it together in order to develop industries performance.

- Lean Tools in Manufacturing and its Importance

Many studies have investigated how the implementation of lean tools can effect on OEE measures, (Shah et al., 2017) have made a study to investigate how the application of lean tools can improve the OEE and the productivity in a heavy fabrication industry, they applied the single-minute exchange of die and cellular layout techniques. The results showed a reduction in the setup time which improved the availability, and with improving the availability of the OEE measure has improved about 6%. and by using cellular layout techniques they improved productivity and so the OEE measure increased by 4.88%.

Another study conducted by (Dadashnejad& Valmohammadi, 2019) showed how the effect of value stream mapping on overall equipment effectiveness measures, and the results showed that the implementation of the suggested improvements through value stream mapping will affect the measures of the OEE positively.

Shakil & Parvez (2020) have also studied the application of value stream mapping (VSM) in a sewing line for improving OEE, their results showed an increased percentage of OEE from 45 to 53.75% when applying the VSM recommended actions into the sewing line.

Applying simulation approach is also a well-known methodology to suggest different improvement scenarios and show the effect of lean tools on the overall performance. Mustafa & Cheng (2017) studied the Improving production changeovers effect on the overall performance using A simulation based virtual process approach, which encompasses the changes in facility layout, process automation, process mapping and changeover cycles in the production system. Their conclusions showed a significant effect on the overall performance.

Another study that used VSM and simulation approach to study the effect on the overall performance is conducted by (Atieh et al., 2016) in which they applied lean manufacturing system on a factory in Jordan to improve the manufacturing process of this industry. They used VSM tool to identify nonvalue added activities, then a discrete event simulation model was employed to show different improvement scenarios. The results of applying some suggested changes were increased performance by 32% and decreased the waiting time by 6%.

- Green Practices in Manufacturing and its Importance

Paul et al. (2014) studied the reasons why green manufacturing is important. Their paper described the use of green manufacturing and its application, and the methods of green manufacturing and they mentioned how applying these methods can improve green image and competitive advantage and how it increases the overall performance in the industry.

Another research paper by Rusinko (2007) suggests that green manufacturing practices may positively affect the competitive outcomes. And they related that some of the green practices such as pollution prevention can outcome different competitive outcomes, for example the manufacturing cost and product quality.

Rehman et al. (2016) have presented an empirical assessment and guides about measuring impact of green manufacturing practices on the organizational performance in Indian industries. their results showed a Positive relationship between these factors, and they worked on develop a model to link between both the critical factors and the performance measures.

- Lean and Green Interrelation

The relationship between those two factors has been investigated and a positive justification has been concluded through all related research papers.

Dües et al. (2013) focus in their study on the relationship between lean and green. Their analysis of the literature has identified the relation between lean and green practices. They found that lean practices are beneficial for green practices, and the implementation of green practices also positively affects businesses performance.

Azevedo et al. (2012) have shown that lean and green practices improve economic, social, and environmental performance and they are related to each other in a positive manner.

Yang et al. (2011) have examined the relationship between lean manufacturing practices, environmental management, and work performance, the result showed a positive relationship between all these factors. The study also showed an important result, which is that applying green practices alone without looking at the lean side may have negative consequences, as it may increase operating costs and lower profit.

Diaz-Elsayed et al. (2013) have discussed how the integration of lean and green practices in an automotive industry can affect the overall performance, the results showed a significant decrease in total costs.

Pampanelli et al. (2014) have developed a model that integrate the environmental sustainability into lean approach and apply this model in a production cell. The result was a reduction in operational costs by 5%-10%.

Prasad & Sharma (2014) indicated in their research that the key to improving the overall performance and sustainability of any organization is to use the similar benefits of lean and green which is eliminating all kinds of wastes, this shows how the two practices are affecting each other in a positive relationship.

Another case study was conducted by (Baysan, et al. ,2019), they studied the effect of applying lean tools on energy efficiency using a simulation -based methodology in a power distribution industry, the results showed a 72.37% reduction in energy consumption.

- Lean and Green Practices to Improve Overall Performance

Sproedt et al. (2015) established a Discrete-event simulation with the use of energy value stream mapping to find the interrelation between the economic and environmental performance dimensions in production systems. They established a model to facilitate the implementation of lean and green practices and implanted it in a real case study and the results showed a positive contribution to developing the overall performance.

Another study in this context was by Diaz-Elsayed et al. (2013) in which they identify an approach for implementing both lean and green strategies into an automotive industry, the combination of lean and green strategies resulted in the reduction of approximately 10.8% of the production costs.

Muñoz-Villamizar et al. (2019) introduced a new tool to integrate, measure and control the productivity and the environmental performance together, the tool called overall greenness performance for value stream mapping (OGP-VSM); after applying it in an automotive company in Spain, it proofed it effectiveness in improving the company overall performance.

Hartini& Ciptomulyono (2015) examined the impact of lean and sustainable manufacturing to improve performance through a comprehensive literature review, they reviewed 58 related papers and found a clear positive relation between these two factors.

Miller& Standridge (2010) has integrated lean tools and sustainability concepts using discrete event simulation modeling and analysis beside mathematical optimization in the furniture industry and it showed a positive impact on the environment and on its own overall performance and financial benefits

The use of OEEE as an evaluation metric for the implantation of green and lean practices has been used only in one case study by Domingo et al. (2015). They have added Sustainability metric to the original OEE equation to become:

$$OEEE = OEE \times Sustainability$$

In their study they have implemented this formula in a real case and the proposed method with the use of lean and green practices and the results showed a reduction in the product cost product by 6.2%.

In this study , I used the methodology suggested by Sproedt et al. (2015) and combined it with OEEE measure suggested by Domingo et al. (2015) in order to evaluate the adoption of lean and green practices in the overall performance starting from evaluating the current state using eVSM and simulation model , then found the areas of improvements and suggest different scenarios , and last thing to evaluate the future scenarios using OEEE as an indicator of improvement.

**Methodology and Results**

In my research, I pioneered the integration of the methodology proposed by Sproedt et al. (2015) with the OEEE metric introduced by Domingo et al. (2015), offering a comprehensive evaluation of lean and green practice adoption. The assessment process encompassed the following key phases:

1- Assessing the current state using eVSM and current OEEE. and a simulation model.
2- Identifying areas for improvement and proposing various improvement scenarios and test them using different simulation models.
3- Evaluating future scenarios using eVSM OEEE as a key performance indicator to measure improvements and present future phase.

This innovative combination of methodologies provided a unique and comprehensive approach that had not been previously explored in literature, offering valuable insights into optimizing operational efficiency while achieving sustainability goals.

Figure 1 shows the steps involved in implementing the project. Consumption of energy has been considered as a metric to evaluate the sustainability of production processes. Figure 2 shows the e-VSM of the current state for the production line considered. The current state has been simulated using Arena simulation software, verification and validation of the simulation module have been also applied, Figure 3 shows the simulation results of the current process, showing the value-added time metric and

The current OEEE has been calculated as below:

$$OEEE = Availability \times Performance \times Quality \times Sustainability$$

**Current OEEE value** = 26 % * 60% * 95% *88 % = **13.1 %**

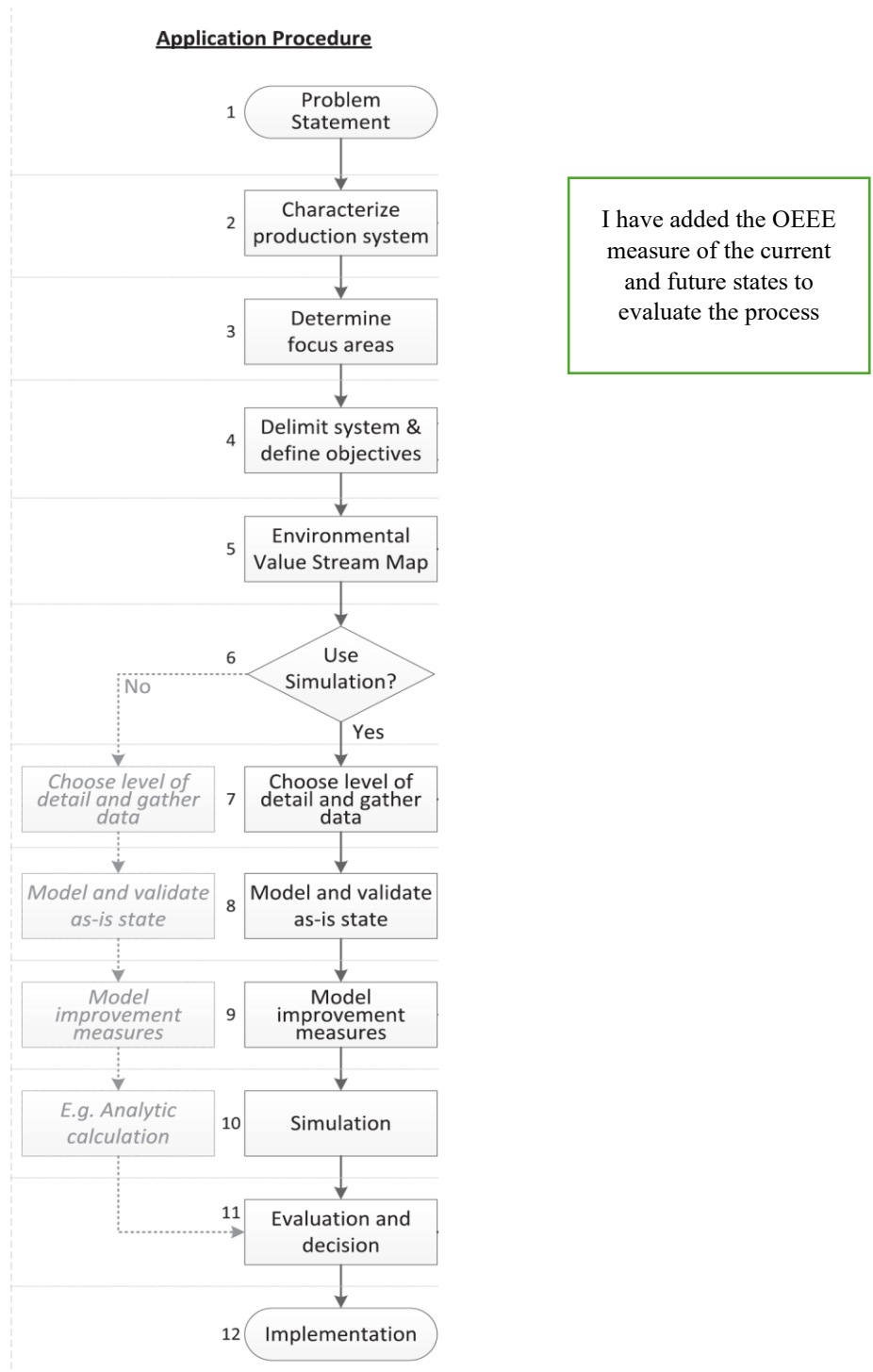

> I have added the OEEE measure of the current and future states to evaluate the process

Figure 1- Research Methodology (Domingo et al., 2015)

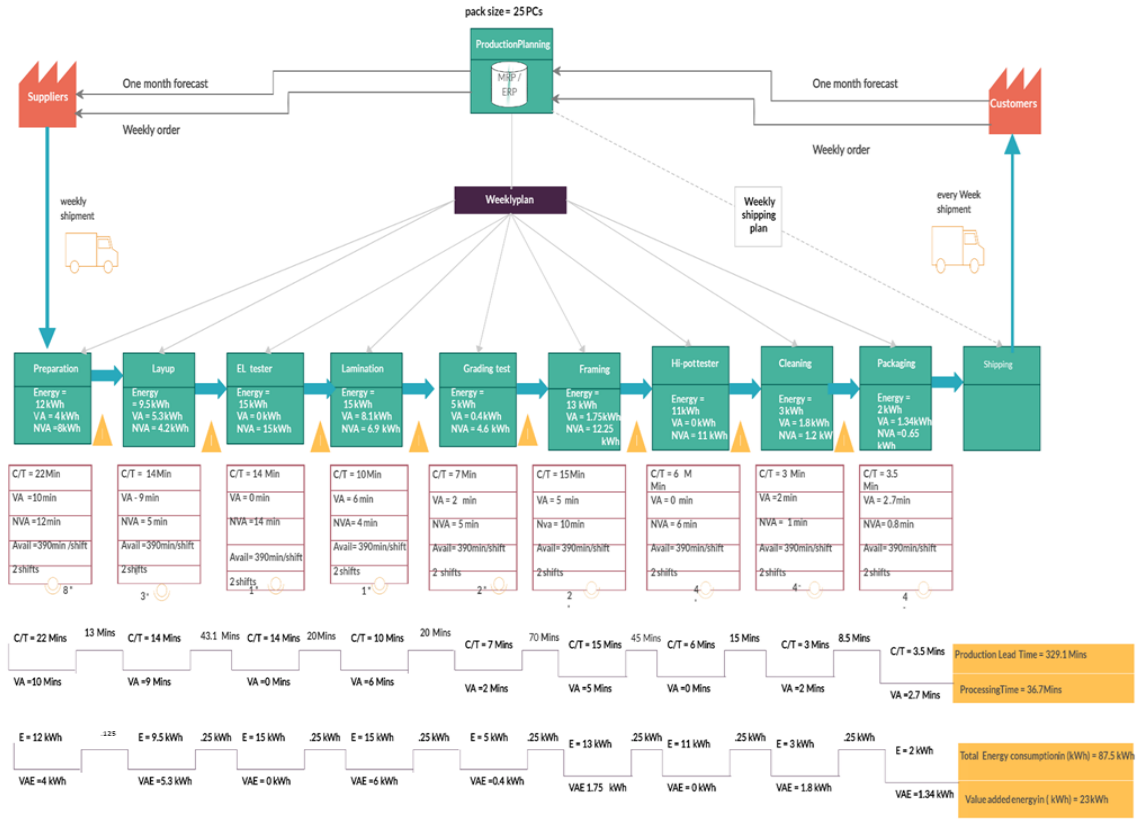

Figure 2- Current Energy Value Stream Mapping

Figure 3- Current Simulation Model Results

The analysis of the current state results to identify wastes and improvements suggestions have been then conducted. The OEEE value indicates that the current production must be improved to increase its value, as we can note from the OEEE value, the performance and sustainability have the lowest values compared to the availability and quality, so our focus should be on improving them to improve the OEEE value. Studies showed that 50% and above is considered acceptable rate, with perfect rate of 100% (Hansen, 2001). For improvements suggestions, process cycle efficiency (PCE (η)) must be calculated to determine which processes have the lowest values and must be improved, based on the following equations for both the energy and time for each workstation (Mehta et al. ,2017):

PCE (η) for time = (T VA / cycle time) * 100 %
PCE (η) for energy = (E VA / total energy) * 100 %

Figure 4 shows the results.

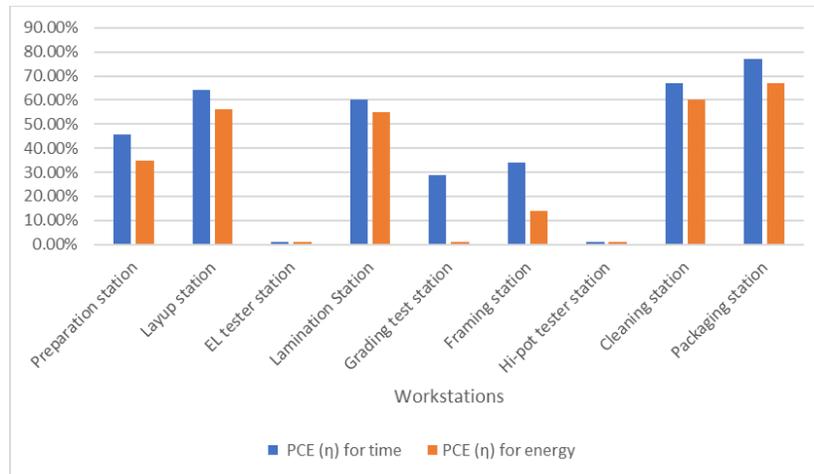

Figure 4 -Current Process Results Summary

After analyzing the data, the following conclusions have been drawn regarding the current state:
1. The significant gap between the takt time (168 minutes per batch) and cycle time (329.1 minutes per batch) indicates inefficiencies in meeting demand.
2. The low OEEE values highlight the need for improvements, with a focus on performance and sustainability.
3. The EL tester and Hi-pot tester have 0% PCE (η) values for time and energy, while the grading tester also shows low values, as inspection stations are non-value-added processes that increase energy consumption and waiting times.
4. The framing workstation ranks second in low PCE (η) values and high waiting times due to the drying process, which takes 6 hours per batch.
5. The preparation workstation has low time efficiency due to cell inspection, high movements, and loading/unloading processes, leading to increased waste.
6. Bottlenecks are observed in the framing, EL tester, and Hi-pot tester workstations, which exhibit the highest waiting times.

7. Lamination and layup stations demonstrate relatively better efficiency, while cleaning and packaging workstations operate optimally with the highest PCE ($\eta$) values and minimal waiting times.

Improvement Suggestions:
1. Optimize Inspection Workstations:
    - Combine or rearrange the three inspection stations and the cell inspection in the preparation stage to reduce redundancy and energy consumption.
    - Merge cell sorting and inspection into a single step.
    - Integrate the EL testing with the layup station and the Hi-pot testing with the framing workstation to eliminate unnecessary workstations.

2. Reduce Drying Time in Framing Workstation:
    - Replace the current 6-hour silicon drying process with UV light/radiation technology, which can reduce drying time to approximately 5 minutes per module, significantly improving production efficiency (Stowe, 1996).

After that, each scenario has been implemented and the improvements areas in the current eVSM have been shown and then I simulate the suggested improvements by changing on the current state simulation model, and after that, calculate cycle time, energy and time efficiency and the OEEE value, using the same tools and equations described above.

The results of each scenario are described below:

1- Scenario 1: Combine, Rearrange the Inspection Workstations
   Figure 5 shows the simulation results using the suggested improvement in the first scenario.

| Category Overview | | | | | | |
|---|---|---|---|---|---|---|
| Values Across All Replications | | | | | | |
| Replications: 30  Time Units: Hours | | | | | | |
| **Entity** | | | | | | |
| **Time** | | | | | | |
| VA Time | Average | Half Width | Minimum Average | Maximum Average | Minimum Value | Maximum Value |
| solar modules | 0.5575 | 0.07 | 0.0936 | 0.9764 | 0.08382326 | 4.6263 |
| NVA Time | Average | Half Width | Minimum Average | Maximum Average | Minimum Value | Maximum Value |
| solar modules | 0.3397 | 0.02 | 0.2179 | 0.4491 | 0.1749 | 1.1205 |
| Wait Time | Average | Half Width | Minimum Average | Maximum Average | Minimum Value | Maximum Value |
| solar modules | 1.5314 | 0.26 | 0.05196577 | 2.6779 | 0.00 | 7.0969 |
| Transfer Time | Average | Half Width | Minimum Average | Maximum Average | Minimum Value | Maximum Value |
| solar modules | 0.1012 | 0.02 | 0.02441637 | 0.2213 | 0.00004408 | 1.1415 |

Figure 5 - Simulation Model Results - Scenario 1

Table 1 Time Improvement - Scenario 1

| Parameter | Scenario 1 values (Min) | Current state values (Min) | Improvement rate % |
|---|---|---|---|
| VA time | 33.45 | 36.7 | 9 % |
| NVA time | 20.38 | 50.34 | 60% |
| Waiting time | 97.8 | 230.55 | 57.58 % |
| Transportation time | 6.6 | 12 | 45% |

Table 1 above shows the difference between the current state simulation results and scenario 1 results and the percentage of time improvements; the energy values improve at the same rate of time since the power is constant as Table 2 shows below.

Table 2 Energy Improvement - Scenario 1

| Parameter | Scenario 1 values (kWh) | Current state values (kWh) | Improvement rates % |
|---|---|---|---|
| VA energy | 20 | 23 | 9% |
| NVA energy | 25.62 | 62.5 | 60% |
| Transportation energy | 0.53 | 1 | 45 % |

**Scenario 1 OEEE value** = 43 % * 90% * 95% *88% = **35 %** compared to the current state with 13.1% OEEE value.

2- Scenario 2: Minimize the Drying Time in the Framing Workstation
The following Figure 6 is the arena model simulation results, showing the value added, non-value added and waiting time after applying the suggested improvements:

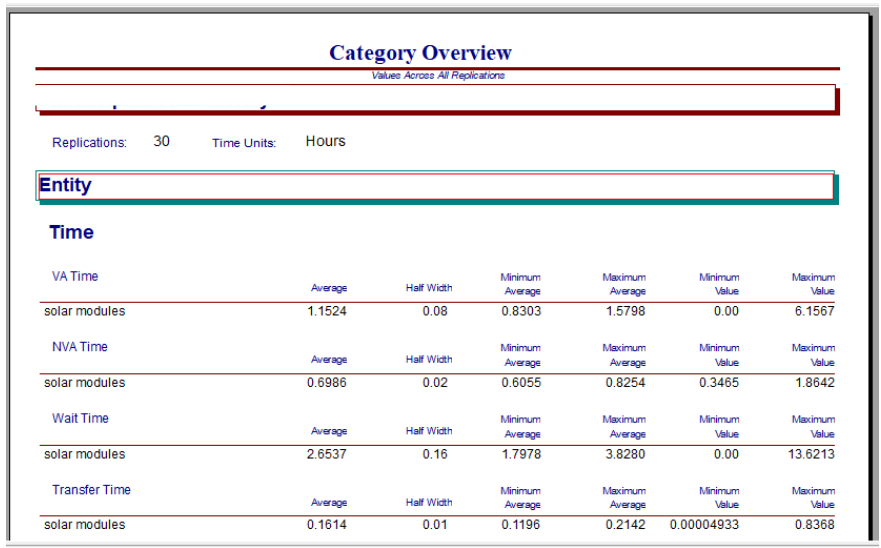

Figure 6 - Simulation Model Results - Scenario 2

Table 3 shows the difference between the current state simulation results and scenario 2 results and the percentage time improvement; the energy values improve at the same rate of time since the power is constant as Table 4 shows.

Table 3 Time Improvement - Scenario 2

| Parameter | Scenario 2 values (Min) | Current state values (Min) | Improvement rate % |
|---|---|---|---|
| VA time | 72 | 36.7 | 44.16% |
| NVA time | 42 | 50.34 | 16.5 % |
| Waiting time | 168 | 230.55 | 27.23% |
| Transfer time | 9.7 | 12 | 20% |

Table 4 Energy Improvement - Scenario 2

| Parameter | Scenario 2 values (kWh) | Current state values (kWh) | Improvement rate % |
|---|---|---|---|
| VA energy | 33 | 23 | 44.16% |
| NVA energy | 52 | 62.5 | 16.5% |
| Transportation energy | 0.8 | 1 | 20% |

**Scenario 2 OEEE value** = 40 % * 95% * 70% *88 % = **24 %** compared to the current state with 13.1 % OEEE value.

**Discussion of the Results**

After applying the suggested improvements, the results showed overall improvement in the production lead time, OEEE value and on the VA time and energy. These results agree with other researchers' findings, as we mentioned before in the literature review, many studies showed a positive impact of applying the lean and green concepts in the overall performance. Combining the research methodology of Sproedt et al. (2015) with OEEE value which suggested by Domingo et al. (2015) allows to test the effect of lean and green practices on the overall performance including quality, availability, performance and sustainability.

In our case, the current state production lead time was 329.1 minutes per batch, which is much greater than takt time which indicates that the customers' orders will expect to be delayed or even canceled besides the need to increase resources by either hiring new workers or paying more to work overtime. The VA time was 36.7 minutes, which is only 12 % of total time, and the waiting time was 230.55 minutes which equates to 70 % of the production lead time, this high percentage caused the presence of bottlenecks as we saw before. The total energy consumption was 86.5 kWh per batch with VA energy of 23 kWh which is also considered low with 27% of total energy. The current OEEE value was 13.1 %, the measures of the OEEE value indicate the existence of problems in the performance and sustainability and suggested that our focus of improvements must make sure to increase these two measures to improve the value of OEEE.

After applying the suggested improvements at the first scenario which suggested to combine and rearrange the inspection workstations based on our findings that these inspection workstations have the lower time and energy efficiency, the results of simulation showed decreasing of the production lead time to become 158.23 minutes which is considered closer to the takt time , and can produce more batches per day ,that predicts that the production rate will become fit with the average customer demand with minimal costs . The VA time increased to equal to 25 % of the total time, the main contribution in lowering the lead time had come from lowering the waiting time which decreased from 230.55 to 97.8 minutes, and that will decrease the bottlenecks. The VA energy has improved to 43 % from total energy consumption. The OEEE value is the main concern in our study since it represents the overall performance was increased to become 35 %, with a major increase in both performance and sustainability.

The other scenario was suggesting to use the UV lighting for drying purposes since the drying process in the framing workstation needs 6 hours per batch, the new method decreased this time to 2.09 hours, this effected immediately the production lead time and deceased to become 292 minutes which is closer to the takt time ,it still need to be improved to reach in average to 168 minutes , but it will minimize the current costs .In this scenario the VA time increase to become 72 minutes which equals to 25 % of the total time ,and the waiting time and NVA time decrease, all of these improvements in time help in improving the value of production lead time, and mentioned before, decreasing the waiting time will help in lowering the bottlenecks .The value-added energy increases to be equal to 40 % from the total energy consumption. The OEEE value became 24 %, which is better than 13.1 % and that indicates an overall improvement in the performance after applying this scenario.

**Conclusions and Recommendations**

The aim of this study was to evaluate the adoption of lean and green practices on the overall performance using simulation model. The energy values and its greenhouse gas equivalencies have been considered as a green metric. After applying the previous scenarios using Arena simulation as a tool to get the future time, and by suggesting different lean and green improvement practices, we have noted that these practices helped in increasing the OEEE values which means that the adoption of lean and green practices have positive impact on the overall performance of the factory. Besides that, the waiting time had been decreased and the value added, and non-value-added time and energy values have been improved, which made the lead time became closer to the takt time, that means that the factory can now satisfy their customer demand with minimal over times and backorders.

The previous improvements have been accomplished by identifying the workstations that need to be improved using the waiting time values and PCE (η) for time and for energy, and by identifying the OEEE measures that must be improved, as we have noted the performance and the sustainability became the main concerns to be improved in order to improve the overall performance since they have the lowest values compared to quality and availability .The values of time and energy efficiency and waiting time, indicated that the inspections workstations have affected the overall performance of the factory, so the first scenario suggested to combine and rearrange these workstations which resulted in increasing the OEEE value from 13.1% to 35 %.

The second scenario suggested using advanced low-cost technology which is the UV light to dry the module at the framing workstation, previously the batch needed 6 hours to get dry, now it only needs 125 minutes to be dried, which affects the overall performance of the factory and increase the OEEE value from 13 % to 24 %. Based on the previous summary, we can note that the first scenario is considered a better option to adopt by the factory since it resulted in a better OEEE and lower production lead time.

The results of this study agree with the previous studies that were mentioned before in the literature review, we all agree that the lean and green practices will have a positive influence on the overall performance of the factory. This study combined the use of simulation models, OEEE and energy value stream mapping to evaluate the performance of the current production line which makes it different from other research. This method of research has been developed from combining a research methodology suggested by Sproedt et al. (2015) with the OEEE value that suggested by Domingo et al. (2015).

- Future work to enhance the study:

    1- Include indirect energy consumption like lighting in calculations.
    2- Apply the suggested method on other Industries to validate the results and conclusions on a larger scale.
    3- Use this methodology to evaluate the adoption of lean and green on other environmental measures rather than energy consumption and its greenhouse gases emissions.

# References

Abreu, M. F., Alves, A. C., & Moreira, F. (2017). Lean-Green models for eco-efficient and sustainable production. *Energy*, *137*, 846-853.

Atieh, A. M., Kaylani, H., Almuhtady, A., & Al-Tamimi, O. (2016). A value stream mapping and simulation hybrid approach: application to glass industry. *The International Journal of Advanced Manufacturing Technology*, *84*, 1573-1586.

Azevedo, S. G., Carvalho, H., Duarte, S., & Cruz-Machado, V. (2012). Influence of green and lean upstream supply chain management practices on business sustainability. *IEEE Transactions on Engineering Management*, *59*(4), 753-765.

Baines, T., Brown, S., Benedettini, O., & Ball, P. D. (2012). Examining green production and its role within the competitive strategy of manufacturers. *Journal of Industrial Engineering and Management*, 53-87.

Baysan, S., Kabadurmus, O., Cevikcan, E., Satoglu, S. I., & Durmusoglu, M. B. (2019). A simulation-based methodology for the analysis of the effect of lean tools on energy efficiency: An application in power distribution industry. *Journal of cleaner production*, *211*, 895-908.

Dadashnejad, A. A., & Valmohammadi, C. (2019). Investigating the effect of value stream mapping on overall equipment effectiveness: a case study. *Total Quality Management & Business Excellence*, *30*(3-4), 466-482.

Diaz-Elsayed, N., Jondral, A., Greinacher, S., Dornfeld, D., & Lanza, G. (2013). Assessment of lean and green strategies by simulation of manufacturing systems in discrete production environments. *CIRP Annals*, *62*(1), 475-478.

Domingo, R., & Aguado, S. (2015). Overall environmental equipment effectiveness as a metric of a lean and green manufacturing system. *Sustainability*, *7*(7), 9031-9047.

Dües, C. M., Tan, K. H., & Lim, M. (2013). Green as the new Lean: how to use Lean practices as a catalyst to greening your supply chain. *Journal of cleaner production*, *40*, 93-100.

Engin, B. E., Martens, M., & Paksoy, T. (2019). Lean and green supply chain management: A comprehensive review. *Lean and Green Supply Chain Management: Optimization Models and Algorithms*, 1-38.

Garza-Reyes, J. A. (2015). Lean and green–a systematic review of the state of the art literature. *Journal of cleaner production*, *102*, 18-29.

Hansen, R. C. (2001). *Overall equipment effectiveness: a powerful production/maintenance tool for increased profits*. Industrial Press Inc..

Hartini, S., & Ciptomulyono, U. (2015). The relationship between lean and sustainable manufacturing on performance: literature review. *Procedia Manufacturing*, *4*, 38-45.

Jiang, R. J., & Bansal, P. (2003). Seeing the need for ISO 14001. *Journal of Management Studies*, *40*(4), 1047-1067.

Maruthi, G. D., & Rashmi, R. (2015). Green Manufacturing: It's Tools and Techniques that can be implemented in Manufacturing Sectors. *Materials Today: Proceedings*, *2*(4-5), 3350-3355.

Miller, G., Pawloski, J., & Standrigde, C. R. (2010). A case study of lean, sustainable manufacturing. *Journal of industrial engineering and management*, *3*(1), 11-32.

Muñoz-Villamizar, A., Santos, J., Garcia-Sabater, J. J., Lleo, A., & Grau, P. (2019). Green value stream-mapping approach to improving productivity and environmental performance. *International Journal of Productivity and Performance Management*, *68*(3), 608-625.

Mustafa, K., & Cheng, K. (2017). Improving production changeovers and the optimization: A simulation based virtual process approach and its application perspectives. *Procedia Manufacturing*, *11*, 2042-2050.

Pampanelli, A. B., Found, P., & Bernardes, A. M. (2014). A Lean & Green Model for a production cell. *Journal of cleaner production*, *85*, 19-30.

Paul, I. D., Bhole, G. P., & Chaudhari, J. R. (2014). A review on green manufacturing: it's important, methodology and its application. *Procedia materials science*, *6*, 1644-1649.

Prasad, S., & Sharma, S. K. (2014). Lean and green manufacturing: concept and its implementation in operations management. *International Journal of Advanced Mechanical Engineering*, *4*(5), 509-514.

Rehman, M. A., Seth, D., & Shrivastava, R. L. (2016). Impact of green manufacturing practices on organisational performance in Indian context: an empirical study. *Journal of cleaner production*, *137*, 427-448.

Rusinko, C. (2007). Green manufacturing: an evaluation of environmentally sustainable manufacturing practices and their impact on competitive outcomes. *IEEE transactions on engineering management*, *54*(3), 445-454.

Shah, M. K., Deshpande, V. A., & Patil, R. M. (2017, February). Case study: Application of Lean tools for Improving overall equipment effectiveness (OEE) & productivity in panel shop of heavy fabrication industry. In *Proceedings of 2nd international conference on emerging trends in mechanical engineering* (pp. 4430-4447).

Shakil, S. I., & Parvez, M. (2020). Application of value stream mapping (VSM) in a sewing line for improving overall equipment effectiveness (OEE): A case study. In *Intelligent Manufacturing and Energy Sustainability: Proceedings of ICIMES 2019* (pp. 249-260). Springer Singapore.

Sproedt, A., Plehn, J., Schönsleben, P., & Herrmann, C. (2015). A simulation-based decision support for eco-efficiency improvements in production systems. *Journal of Cleaner Production*, *105*, 389-405.

Stowe, R. W. (1996, November). High-power UV lamps for industrial UV curing applications. In *Ultraviolet Atmospheric and Space Remote Sensing: Methods and Instrumentation* (Vol. 2831, pp. 208-219). SPIE.

Thanki, S., Govindan, K., & Thakkar, J. (2016). An investigation on lean-green implementation practices in Indian SMEs using analytical hierarchy process (AHP) approach. *Journal of cleaner production*, *135*, 284-298.

Yang, M. G. M., Hong, P., & Modi, S. B. (2011). Impact of lean manufacturing and environmental management on business performance: An empirical study of manufacturing firms. *International Journal of production economics*, *129*(2), 251-261.